\documentclass{emulateapj}

\newcommand{\Kepler}{{\it Kepler}}

\newcommand{\hipparcos}{{\it Hipparcos}}
\newcommand{\vespa}{\texttt{vespa}}

\newcommand{\be}{\begin{equation}}
\newcommand{\ee}{\end{equation}}

\newcommand{\metallicity}{[M/H]}
\newcommand{\msun}{M$_\odot$}
\newcommand{\rsun}{R$_\odot$}

\newcommand{\kms}{\ensuremath{\rm km\,s^{-1}}}
\newcommand{\ms}{\ensuremath{\rm m\,s^{-1}}}

\newcommand{\teff}{5370}
\newcommand{\teffe}{50}

\newcommand{\logg}{4.52}
\newcommand{\logge}{0.03}

\newcommand{\mh}{0.00}
\newcommand{\mhe}{0.08}

\newcommand{\mstar}{0.883}
\newcommand{\mstare}{0.033}

\newcommand{\rstar}{0.828}
\newcommand{\rstare}{0.028}

\newcommand{\mearth}{M$_\oplus$}
\newcommand{\rearth}{R$_\oplus$}

\newcommand{\ldone}{0.27}
\newcommand{\uldone}{0.12}
\newcommand{\ldtwo}{0.62}
\newcommand{\uldtwo}{0.21}
\newcommand{\rprstb}{0.01725}
\newcommand{\urprstb}{0.00023}
\newcommand{\arstb}{4.56}
\newcommand{\uarstb}{^{+0.079}_{-0.11}}
\newcommand{\inclb}{88.4}
\newcommand{\uinclb}{^{+1.1}_{-1.5}}
\newcommand{\impb}{0.131}
\newcommand{\uimpb}{0.098}
\newcommand{\rplb}{1.557}
\newcommand{\urplb}{0.057}
\newcommand{\perplb}{0.959628}
\newcommand{\uperplb}{0.000012}
\newcommand{\ttransitb}{2457394.37450}
\newcommand{\uttransitb}{0.00044}
\newcommand{\rprstc}{0.0314}
\newcommand{\urprstc}{^{+0.0028}_{-0.0011}}
\newcommand{\arstc}{42.2}
\newcommand{\uarstc}{^{+6.7}_{-13}}
\newcommand{\inclc}{89.27}
\newcommand{\uinclc}{^{+0.51}_{-0.85}}
\newcommand{\impc}{0.54}
\newcommand{\uimpc}{0.26}
\newcommand{\rplc}{2.85}
\newcommand{\urplc}{^{+0.24}_{-0.15}}
\newcommand{\perplc}{29.8454}
\newcommand{\uperplc}{0.0012}
\newcommand{\ttransitc}{2457394.9788}
\newcommand{\uttransitc}{0.0012}

\newcommand{\thisstar}{HD~3167}
\newcommand{\thisfirstplanet}{HD~3167~b}
\newcommand{\thissecondplanet}{HD~3167~c}

\usepackage{xcolor}
\usepackage{hyperref}
\usepackage{breakurl}
\usepackage{amsmath}
\usepackage{graphicx}
\usepackage{verbatim}
\usepackage{booktabs}

\slugcomment{}

\shorttitle{Two Small Planets Transiting HD 3167}
\shortauthors{Vanderburg et al.}

\begin{document}


\title{Two Small Planets Transiting HD 3167}
\author{Andrew Vanderburg\altaffilmark{1,7,8}, Allyson Bieryla\altaffilmark{1}, Dmitry A. Duev\altaffilmark{2}, Rebecca Jensen-Clem\altaffilmark{2}, David W. Latham\altaffilmark{1}, Andrew W. Mayo\altaffilmark{1}, Christoph Baranec\altaffilmark{3}, Perry Berlind\altaffilmark{1}, Shrinivas Kulkarni\altaffilmark{2},  Nicholas M. Law\altaffilmark{4}, Megan N. Nieberding\altaffilmark{5}, Reed Riddle\altaffilmark{2}, \& Ma\"{i}ssa Salama\altaffilmark{6}}

\altaffiltext{1}{Harvard--Smithsonian Center for Astrophysics, 60 Garden St., Cambridge, MA 02138, USA}
\altaffiltext{2}{California Institute of Technology, Pasadena, CA, 91125, USA}

\altaffiltext{3}{University of Hawai`i at M\={a}noa, Hilo, HI 96720, USA}
\altaffiltext{4}{University of North Carolina at Chapel Hill, Chapel Hill, NC, 27599, USA}
\altaffiltext{5}{National Optical Astronomy Observatory, 950 N. Cherry Avenue, Tucson, AZ 85719, USA}
\altaffiltext{6}{University of Hawai`i at M\={a}noa, Honolulu, HI 96822, USA}
\altaffiltext{7}{NSF Graduate Research Fellow}
\altaffiltext{8}{\url{avanderburg@cfa.harvard.edu}}


\begin{abstract}
We report the discovery of two super-Earth-sized planets transiting the bright (V = 8.94, K = 7.07) nearby late G-dwarf \thisstar, using data collected by the K2 mission. The inner planet, \thisfirstplanet, has a radius of 1.6 \rearth\ and an ultra-short orbital period of only 0.96 days. The outer planet, \thissecondplanet, has a radius of 2.9 \rearth\ and orbits its host star every 29.85 days. At a distance of just 45.8 $\pm$ 2.2 pc, \thisstar\ is one of the closest and brightest stars hosting multiple transiting planets, making HD 3167 b and c well suited for follow-up observations. The star is chromospherically inactive with low rotational line-broadening, ideal for radial velocity observations to measure the planets' masses. The outer planet is large enough that it likely has a thick gaseous envelope which could be studied via transmission spectroscopy. Planets transiting bright, nearby stars like \thisstar\ are valuable objects to study leading up to the launch of the James Webb Space Telescope.

\end{abstract}

\keywords{planetary systems --- planets and satellites: detection --- stars: individual (HD 3167)}

\section{Introduction}

Transiting exoplanets are benchmark objects. Like an eclipsing binary star, a transiting exoplanet offers unique opportunities for a rich variety of follow-up studies due to its favorable orbital geometry. It is possible to measure a planet's fundamental bulk properties like mass and radius \citep[e.g.][]{dai, gettel}, study its atmosphere photometrically or spectroscopically \citep[][]{diamondlowe, knutson, kreidberg}, and measure the alignment between the planet's orbit and the host star's spin axis \citep[e.g.][]{sanchisojeda}. 

More so than eclipsing binary stars, transiting exoplanets are difficult to detect. Wide-field ground-based transit surveys have detected hundreds of hot Jupiters \citep[e.g.][]{wasps}, but the sensitivity of these surveys falls off quickly at longer orbital periods \citep{gaudiperiod} and smaller planet radii \citep{gaudiradius}. Space telescopes like NASA's \Kepler\ observatory \citep{koch}, with smaller fields of view but better photometric precision have been highly successful at detecting small planets \citep{coughlin,morton16}, but these planets typically orbit faint stars due to the narrow survey design. \Kepler\ revolutionized our knowledge of exoplanets, but only a handful of \Kepler's discoveries orbit stars bright enough for detailed follow-up observations. In its extended K2 mission, \Kepler\ is surveying a larger area of sky and is finding more exciting planets suitable for follow-up observations \citep{crossfield}, but planets transiting stars brighter than 9th magnitude remain a rare prize. Pencil-beam ground-based transit surveys like MEarth \citep{irwin}, APACHE \citep{sozzetti}, and TRAPPIST \citep{trappist} have found what are likely among the best planets in the sky for atmospheric characterization, but the number of transiting planets detected by those surveys to date is small enough to count on one hand \citep{charbonneau,gj1132,trappist1}. Today, planets transiting the brightest known host stars in the sky were uncovered by years or decades of precise radial velocity (RV) measurements \citep{winn55cnc, dragomir, motalebi}. While RV searches have been fruitful, the observations are challenging, the surveys require many nights over many years on large telescopes, and the success rate of finding transiting planets is low. 
\begin{figure*}[ht!] 
   \centering
   \includegraphics[width=6.5in]{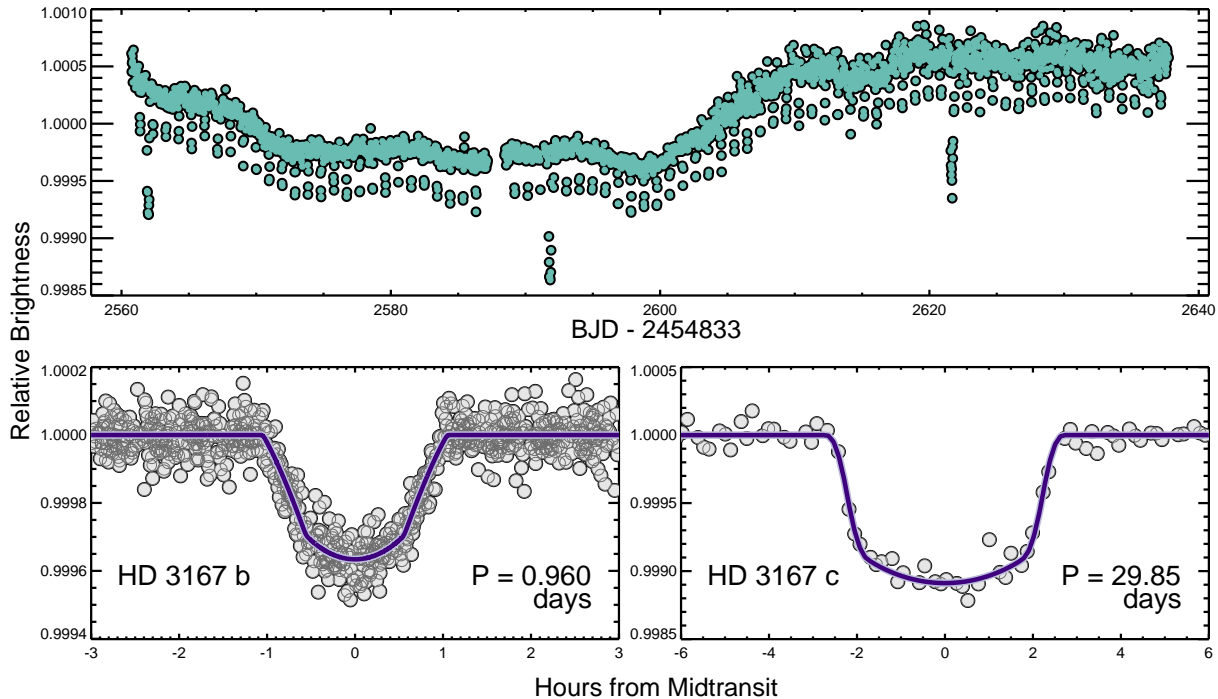} 
   \caption{K2 light curve of \thisstar. Top: the full K2 light curve. Both the numerous, shallow transits of \thisfirstplanet\ and three deeper transits of \thissecondplanet\ are evident in the light curve by eye. Bottom left: K2 light curve (grey dots) phase folded on the transits of \thisfirstplanet, and best-fit transit model (thick purple line). Bottom right: K2 light curve (grey dots) phase folded on the transits of \thissecondplanet, and best-fit transit model (thick purple line).}
   \label{lc}
\end{figure*}

NASA's TESS mission \citep{ricker}, an all-sky space-based transit survey, is expected to address the need for planets transiting bright host stars by discovering hundreds of the nearest and brightest transiting exoplanets in the sky \citep{sullivan}, but TESS won't begin collecting data until early 2018. There is an immediate need for small planets transiting bright stars that can be studied before the launch of missions like TESS and the James Webb Space Telescope (JWST). The lessons learned from these first planets to be examined in detail will help inform decisions about how to use resources like JWST to efficiently learn about exoplanets.

Here, we announce the discovery of two super-Earth-sized planets transiting the bright (V = 8.94, K = 7.07) nearby dwarf star \thisstar\ using data from the K2 mission. The inner planet, \thisfirstplanet, is a 1.6 \rearth\ super-Earth that orbits its host every 0.96 days. \thisfirstplanet\ has likely lost most of any atmosphere it once possessed due to the intense radiation environment in its short-period orbit \citep{robertousp}. The outer planet, \thissecondplanet, orbits in a 29.85 day period, and with a radius of 2.9 \rearth\ likely has a thick gaseous envelope. The host star is both bright enough in visible wavelengths for precise RV follow-up to measure the planets' masses, and bright enough in infrared wavelengths to spectroscopically interrogate \thissecondplanet's atmosphere. In Section \ref{observations} we describe our observations of \thisstar, our data reduction, and our analysis. In Section \ref{validation}, we describe our statistical validation of the planet candidates, and in Section \ref{discussion}, we discuss the \thisstar\ planets and their importance in the context of the thousands of known transiting exoplanets.

\section{Observations and Analysis}\label{observations}

\subsection{K2 Light Curve}\label{lightcurve}

\Kepler\ observed \thisstar\ between 3 January 2016 and 23 March 2016 during Campaign 8 of its K2 mission. We identified two planet candidates transiting \thisstar\ after processing pixel level data to produce a light curve, removing systematic effects due to \Kepler's unstable pointing \citep{vj14}, and searching for planets using a Box Least Squares periodogram search \citep{kovacs, v15}. We extracted the light curve from a photometric aperture shaped like the image of the star on the detector, with an average radius of about 20\arcsec. After identifying the transits, we reprocessed the raw K2 light curve to remove systematics while simultaneously fitting for stellar variability and the two planets' transit signals \citep{v15}. The noise level in the first half of the K2 light curve is a \Kepler-like 10 ppm per six hours, but the second half of the light curve is about 80\% noisier, likely due to \Kepler's more erratic motion during that part of the campaign when the median point-to-point motion increased from 0.13\arcsec to 0.29\arcsec. The full K2 light curve is shown in Figure \ref{lc}. 

We measured transit parameters by fitting the two transit signals in the K2 light curve with \citet{mandelagol} models using a Markov Chain Monte Carlo algorithm with an affine invariant ensemble sampler \citep{goodman}. Our model has 14 free parameters: two quadratic limb darkening coefficients \citep[using the $q_1$ and $q_2$ parametrization from][]{kippingld}, the uncertainty of each photometric data point in the first and second halves of the campaign, and for each planet the orbital period, transit time, cosine of orbital inclination ($\cos{i}$), planet to star radius ratio ($R_p/R_\star$), and scaled semimajor axis ($a/R_\star$). We assumed a circular orbit for the inner planet, and imposed a prior on the mean stellar density (from the stellar mass and radius calculated in Section \ref{spectroscopy}) to constrain the scaled semimajor axis. We accounted for \Kepler's 29.4 minute long-cadence integration time by oversampling the model light curve by a factor of 30 and performing a trapezoidal integration. We sampled the parameter space with 150 walkers evolved over 50,000 steps, and discarded the first 40,000 steps as ``burn-in.'' A calculation of the Gelman-Rubin potential scale reduction factors for each parameter (all below 1.2) confirmed the MCMC fit had converged. The transit light curves are shown in the bottom panels of Figure \ref{lc} along with the best-fit model. Transit parameters and uncertainties are listed in Table \ref{bigtable}.

\subsection{Spectroscopy}\label{spectroscopy}

We observed \thisstar\ with the Tillinghast Reflector Echelle Spectrograph (TRES) on the 1.5 m telescope at the Fred L. Whipple Observatory on Mt. Hopkins, AZ. We acquired one spectrum with a spectral resolving power of $\Delta\lambda/\lambda$ = 44,000 on 11 July 2016. A four minute exposure yielded a spectrum with a signal-to-noise ratio of 71 per resolution element at 550 nm. We cross-correlated the spectrum with a number of templates from our library of synthetic spectra, and measured an absolute RV of $19.5~ \pm ~0.1$ \kms\ from the best matched template. The cross correlation function showed no evidence for a second set of spectral lines, indicating there are no nearby, bright blended sources within one or two arcseconds. 

We measured spectroscopic parameters of \thisstar\ using the Stellar Parameter Classification tool \citep[SPC, ][]{buchhave, buchhave14}, which works by cross-correlating observed spectra with a grid of synthetic model spectra generated from \citet{kurucz} model atmospheres. The SPC analysis revealed that \thisstar\ has a temperature of $T_{\rm eff} ~=~  5370~ \pm ~50$ K, metallicity of [M/H] = $0.00~ \pm  ~0.08$, surface gravity of $\log{g_{\rm cgs, SPC}}~ =~ 4.54~ \pm ~0.10$, and has a low projected rotational velocity with only an upper limit on $v\sin{i}~<~2~\kms$. We also measured the flux near the cores of the calcium II H and K lines and calculated Mt. Wilson $S_{\rm HK} = 0.178~ \pm ~0.005$,  and $\log{R'_{\rm HK}} = -4.97~ \pm ~0.02$. The star is evidently chromospherically inactive. 

We estimated fundamental stellar parameters using an online interface\footnote{\url{http://stev.oapd.inaf.it/cgi-bin/param}} which, following \citet{dasilva}, interpolates the star's V-magnitude, parallax, metallicity and effective temperature onto PARSEC model isochrones \citep{bressan}, assuming the lognormal initial mass function from \citet{chabrier}. \thisstar\ is a dwarf star slightly smaller than the Sun, with a mass of \mstar\ $\pm$ \mstare\ \msun\ and a radius of \rstar\ $\pm$ \rstare\ \rsun. The models predict a surface gravity of $\log{g_{\rm cgs}} = 4.52 \pm 0.032$, consistent with the spectroscopic measurement from SPC, indicating that our stellar parameters are sensible. Stellar parameters are listed in Table \ref{bigtable}.

\subsection{Imaging}

We observed \thisstar\ with the Robo-AO adaptive optics system \citep{baranec, lawroboao} installed at the 2.1 m telescope at Kitt Peak National Observatory \citep{roboaokp}. We observed \thisstar\ with an {\emph i'}-band filter on 11 July 2016, taking images at a rate of 8.6 Hz for a total of 120 seconds. In post-processing, we shifted and added the images together using \thisstar\ as a tip/tilt guide star. The resulting image showed no evidence for any stars other than \thisstar. We estimated a contrast curve for the Robo-AO image by measuring the residuals from resolution element-sized regions in the PSF-subtracted image, as described by \citet{roboaokp} and Jensen-Clem et al. (2016, in prep). The Robo-AO image excludes 5-sigma detections of stars 2 magnitudes fainter than \thisstar\ at 0$\farcs$25 and stars 5 magnitudes fainter at 1$\arcsec$. The Robo-AO field of view (36\arcsec$\times$36\arcsec) does not cover the full photometric aperture, so we also inspected archival images from the Palomar Observatory Sky Survey to confirm there is no evidence for faint stars in the photometric aperture at wide separations.

\section{Statistical Validation}
\label{validation}

We validated the planetary nature of the candidates transiting \thisstar\ using statistical techniques developed by \citet{morton2} as implemented in the \vespa\ software package \citep{morton2015}. \texttt{Vespa} calculates the false positive probability (FPP) of transiting planet candidates given knowledge about the location of the planets in the sky (and hence the prevalence of potential background objects which could cause astrophysical false-positives) as well as observational constraints. We calculated FPPs of \thisstar\ b and c taking into account constraints on the depth of a potential secondary eclipse in the K2 light curve, the (lack of) difference in depth between even and odd transits, and the constraints on fainter stars in the photometric aperture from archival imaging and Robo-AO. We constrained the depth of possible secondary eclipses by measuring the scatter in the depths of putative eclipses at different phases in the planet's orbit, and conservatively chose limits of 12 ppm for the inner planet, and 40 ppm for the outer planet. We calculate FPPs of roughly $10^{-3}$ and $10^{-4}$ for \thisfirstplanet\ and \thissecondplanet, respectively. The FPPs for both candidates are low because K2 field 8 is at a high galactic latitude where the density of stars is low so there are few potential background contaminants. The fact that these candidates are found in a multi-planet system further lowers the FPPs by about a factor of 30, which we estimated using the number of single and multi-candidate systems we detected in our search of K2 Campaign 8, following \citet{lissauer}. We therefore consider the candidates transiting \thisstar\ to be validated as genuine exoplanets. 

\section{Discussion}
\label{discussion}

The main importance of the \thisstar\ planetary system is due to the brightness and proximity of the host star. With a V-magnitude of 8.94, slow projected rotation of less than 2 \kms, and low activity, \thisstar\ is highly suitable for precise RV observations to measure the planets' masses. If \thisfirstplanet\ is rocky with a mass of about 4 \mearth, it should induce RV variations with a semiamplitude of about 3 \ms. Depending on its composition, \thissecondplanet\ could induce RV variations with a semiamplitude of anywhere between 1 \ms (for a roughly 5 \mearth\ planet) and 3 \ms (for a roughly 15 \mearth\ planet). These signals should be readily detectable with modern spectrographs.

\thissecondplanet\ is one of the best currently known small planets for atmospheric characterization with transit transmission spectroscopy. We downloaded a list of transiting planets with radii smaller than 4 \rearth\ from the NASA Exoplanet Archive \citep{akeson} and calculated the expected signal-to-noise ($S/N$) one could hope to accumulate per transit compared to the expected scale height of each planets' atmosphere. In particular, we calculated: 

\begin{equation}
S/N \propto \frac{R_{p} H \sqrt{Ft_{14}}}{R_{\star}^2}
\end{equation}
\begin{equation}
H = \frac{k_{b}T_{eq}}{\mu g}
\end{equation}

\noindent where $R_p$ is the planet's radius, $R_\star$ is the star's radius, $H$ is the atmosphere's scale height, $k_{b}$ is Boltzmann's constant, $T_{eq}$ is the planet's equilibrium temperature, $\mu$ is the atmosphere's mean molecular weight, $g$ is the planets' surface gravity, $t_{14}$ is the transit duration, and $F$ is the flux from the star. We calculated $F$ from the host stars' H-band magnitudes to test suitability for observations with the Hubble Space Telescope's Wide Field Camera 3 instrument, and we assumed that planets less dense than rocky structural models \citep{zeng} have atmospheres dominated by molecular hydrogen, while planets consistent with rocky structural models have atmospheres dominated by heavier molecules (like oxygen).

\begin{deluxetable}{llccl}
\tablewidth{0pt}
\tablehead{
\colhead{ } & \colhead{Planet} & \colhead{$R_{\rm P}$ (\rearth)}& \colhead{Predicted $S/N ^{a}$} &\colhead{Discovery}}
\startdata
 1. & GJ 1214 b  &$2.85 \pm 0.20$ & 3.9$^{c}$ & MEarth \\
 2. &GJ 3470 b  &$3.88 \pm 0.32$ & 1.9$^{c}$ & RV \\
 3. &55 Cnc e  & $1.91 \pm 0.08$ & 1.6$^{~}$ & RV \\
 4. &HD 97658 b  & $2.25 \pm 0.10$& 1.1$^{c}$ & RV  \\
 5. &\thissecondplanet\ & $\rplc \pm \urplc$& 1.0$^{b}$ & K2 \\
\enddata
\tablecomments{$a$: the signal-to-noise ratios for transmission spectroscopy per transit are given relative to \thissecondplanet. We note that the predicted signal-to-noise ratios for the five planets listed here are all calculated assuming low mean molecular weight atmospheres. $b$: planet mass used in the $S/N$ calculation estimated using the relation given by \citet{weissmarcy}. $c$: the transmission spectra of these planets are flat, indicating either obscuring clouds/haze layers, or atmospheres with high molecular weights.}
\label{best}
\end{deluxetable}

Table \ref{best} ranks the small transiting exoplanets most amenable to atmospheric characterization. If \thissecondplanet\ has a thick gaseous envelope, as expected based on mass measurements of similarly sized exoplanets \citep{weissmarcy}, only four known small planets are more amenable to atmospheric characterization.  This doesn't necessarily mean atmospheric features will be detected in \thissecondplanet\ -- GJ 1214 b is by far the most amenable small planet for transit spectroscopy, but its transmission spectrum is masked by clouds or hazes \citep{kreidberg}. Indeed, three of the four small planets more amenable to atmospheric characterization than \thissecondplanet\ have flat transmission spectra inconsistent with a hydrogen-dominated atmosphere. A major goal for transmission spectroscopists is understanding which planets form thick clouds or hazes, and on which planets clear skies permit transit spectroscopy. When TESS launches, it will likely find about 80 planets comparable to or better than \thissecondplanet\ for transmission spectroscopy (as calculated above for existing planets). Studying \thissecondplanet\ now could inform the choice of which TESS planets should be observed to most efficiently learn about the atmospheres of small planets.

Unlike most known multi-transiting systems, the planets in the \thisstar\ system are widely separated in orbital period. The period ratio of $P_c/P_b$ = 31.1 is larger than the period ratios of 99\% of all pairs of adjacent planets in the \Kepler\ Data Release 24 planet candidate catalog \citep{coughlin}. This could suggest the presence of additional, non-transiting, planets in the \thisstar\ system which might be revealed by RV observations. 

With a period of just 0.96 days, \thisfirstplanet\ is an example of an ultra-short period (USP) planet, as defined by \citet{robertousp} \citep[although its period is longer than the 12 hour cutoff chosen by][]{jackson}. \citet{robertousp} found that in the \Kepler\ sample, essentially all USP planets have radii smaller than 2 \rearth, indicating that the intense radiation so close to their host stars has stripped the planets of thick gaseous envelopes. Even though RV studies have shown that 1.6 \rearth\ planets often have thick gaseous envelopes, \thisfirstplanet's radiation environment makes it likely its composition is also predominantly rocky. 

Finally, we note that the short period of \thisfirstplanet\ makes it likely that spectroscopic observations of \thissecondplanet's atmosphere might overlap with a transit of the inner planet \citep[see, for example,][]{dewit}. This could be an efficient way to rule out a hydrogen-dominated atmosphere for \thisfirstplanet. Observers should be cautious, however, to ensure that a transit of \thisfirstplanet\ not interfere with out-of-transit observations necessary for calibration.

\acknowledgments
We thank Hannah Diamond-Lowe, Christophe Lovis, and Phil Muirhead for valuable conversations and we thank the anonymous referee for a helpful comments. A.V. is supported by the NSF Graduate Research Fellowship, Grant No. DGE 1144152. D.W.L. acknowledges partial support from the Kepler mission under NASA Cooperative Agreement NNX13AB58A with the Smithsonian Astrophysical Observatory. C.B. acknowledges support from the Alfred P. Sloan Foundation. 

This research has made use of NASA's Astrophysics Data System and the NASA Exoplanet Archive, which is operated by the California Institute of Technology, under contract with the National Aeronautics and Space Administration under the Exoplanet Exploration Program. The National Geographic Society--Palomar Observatory Sky Atlas (POSS-I) was made by the California Institute of Technology with grants from the National Geographic Society. The Oschin Schmidt Telescope is operated by the California Institute of Technology and Palomar Observatory.

This paper includes data collected by the \Kepler\ mission. Funding for the \Kepler\ mission is provided by the NASA Science Mission directorate. Some of the data presented in this paper were obtained from the Mikulski Archive for Space Telescopes (MAST). STScI is operated by the Association of Universities for Research in Astronomy, Inc., under NASA contract NAS5--26555. Support for MAST for non--HST data is provided by the NASA Office of Space Science via grant NNX13AC07G and by other grants and contracts.

Robo-AO KP is a partnership between the California Institute of Technology, University of Hawai`i M\={a}noa, University of North Carolina, Chapel Hill, the Inter-University Centre for Astronomy and Astrophysics, and the National Central University, Taiwan.  Robo-AO KP was supported by a grant from Sudha Murty, Narayan Murthy, and Rohan Murty.  The Robo-AO instrument was developed with support from the National Science Foundation under grants AST-0906060, AST-0960343, and AST-1207891, the Mt. Cuba Astronomical Foundation, and by a gift from Samuel Oschin.  Based in part on observations at Kitt Peak National Observatory, National Optical Astronomy Observatory (NOAO Prop. ID: 15B-3001), which is operated by the Association of Universities for Research in Astronomy (AURA) under cooperative agreement with the National Science Foundation. 

Facilities: \facility{Kepler/K2, FLWO:1.5m (TRES), KPNO:2.1m (Robo-AO)}


\begin{deluxetable*}{lcccc}
\tablecaption{System Parameters for \thisstar \label{bigtable}}
\tablewidth{0pt}
\tablehead{
  \colhead{Parameter} & 
  \colhead{Value}     &
  \colhead{} &
  \colhead{68.3\% Confidence}     &
  \colhead{Comment}   \\
  \colhead{} & 
  \colhead{}     &
  \colhead{} &
  \colhead{Interval Width}     &
  \colhead{}  
}
\startdata
\emph{Other Designations} & & & \\
EPIC 220383386  & & & \\
HIP 2736  & & & \\
\\
\emph{Basic Information} & & & \\
Right Ascension & 00:34:57.52 & & & A \\
Declination & +04 22 53.3 & & & A \\
Proper Motion in RA [\ensuremath{\rm mas\,yr^{-1}}]& 107.6 & $\pm$ & 1.0&A  \\
Proper Motion in Dec [\ensuremath{\rm mas\,yr^{-1}}]& -173.1 & $\pm$ & 0.7&A  \\
Absolute Radial Velocity [\kms]& 19.5 & $\pm$ & 0.1& B \\
Distance to Star~[pc]& 45.8 &$\pm$&2.2& A\\
V-magnitude & 8.94 &$\pm$  & 0.02 & A\\ 
K-magnitude & 7.07 &$\pm$  & 0.02 & A\\ 
\\
\emph{Stellar Parameters} & & & \\
$M_\star$~[$M_\odot$] & \mstar & $\pm$&$ \mstare$ & C \\
$R_\star$~[$R_\odot$] & \rstar & $\pm$&$ \rstare$ & C \\
Limb darkening $q_1$~ & \ldone  & $\pm$&$ \uldone$ & D \\
Limb darkening $q_2$~ & \ldtwo  & $\pm$&$ \uldtwo$ & D \\
$\log g_\star$~[cgs] & \logg & $\pm$& \logge & C \\
Metallicity \metallicity & \mh & $\pm$&\mhe & B \\
$T_{\rm eff}$ [K] & \teff & $\pm$&$ \teffe$ & B\\
$v\sin{i}$ [\kms] &  $< 2$ & & & B\\
Mt. Wilson $S_{\rm HK}$ & 0.178 & $\pm$&$ 0.005$ & B\\
Mt. Wilson $\log{R'_{\rm HK}}$ & -4.97 & $\pm$&$ 0.02$ & B\\

 & & \\
 
\emph{\thisfirstplanet} & & & \\
Orbital Period, $P$~[days] & \perplb & $\pm$&$ \uperplb $ & D \\
Radius Ratio, $R_P/R_\star$ & \rprstb & $\pm$ &$ \urprstb$ & D \\
Scaled semimajor axis, $a/R_\star$  & \arstb & &$ \uarstb$ & D \\
Orbital inclination, $i$~[deg] & \inclb & &$ \uinclb$ & D \\
Transit impact parameter, $b$ & \impb & $\pm$&$ \uimpb$ & D \\
Transit Duration, $t_{14}$~[hours] & 1.65 &$\pm$ &$0.029$ & D \\
Time of Transit $t_{t}$~[BJD] & \ttransitb & $\pm$& \uttransitb & D\\ 
$R_P$~[\rearth] & \rplb &   $\pm$&$ \urplb$  & C,D \\
$T_{eq} = T_{\rm eff}(1 - \alpha)^{1/4}2^{-1/4}\sqrt{\frac{R_\star}{a}}$~[K] & 1860 & $\pm$  & 160  & B,C,D,E \\
 & & \\
 
\emph{\thissecondplanet} & & & \\
Orbital Period, $P$~[days] & \perplc & $\pm$&$ \uperplc $ & D \\
Radius Ratio, $R_P/R_\star$ & \rprstc & &$ \urprstc$ & D \\
Scaled semimajor axis, $a/R_\star$  & \arstc & &$ \uarstc$ & D \\
Orbital Inclination, $i$~[deg] & \inclc & &$ \uinclc$ & D \\
Transit Impact parameter, $b$ & \impc & $\pm$&$ \uimpc$ & D \\
Transit Duration, $t_{14}$~[hours] & 4.81 & &$^{+0.20}_{-0.11}$ & D \\
Time of Transit $t_{t}$~[BJD] & \ttransitc & $\pm$& \uttransitc & D\\ 
$R_P$~[\rearth] & \rplc &   &$ \urplc$  & C,D \\
$T_{eq} = T_{\rm eff}(1 - \alpha)^{1/4}\sqrt{\frac{R_\star}{2a}}$~[K] & 500 & $\pm$  & 40  & B,C,D,E \\

 & & \\

\enddata

\tablecomments{A: Parameters come from the EPIC catalog \citep{epic}. B: Parameters come from analysis of the TRES spectrum. C: Parameters come from interpolation of the \hipparcos\ parallax \citep{hipparcos}, V-magnitude, metallicity, and effective temperature onto model isochrones. D: Parameters come from analysis of K2 light curve, using samples from the MCMC fit described in Section \ref{lightcurve}. E: Equilibrium temperatures $T_{eq}$ calculated assuming circular orbits, and an albedo $\alpha$ uniformly distributed between 0 and 0.7. We report the day-side temperature for \thisfirstplanet\ and the average temperature for \thissecondplanet\ assuming perfect heat redistribution.}

\end{deluxetable*}
\clearpage

\end{document}